\documentstyle[aps,prl,twocolumn,epsf]{revtex}
\begin{document}
\noindent
{\bf Newman and Swift Reply:} \ We agree with the authors of the
previous Comment\cite{pc} that the percolative effect which occurs at
$\alpha = -1$ (for $d>1$) offers a very useful insight into
the behaviour of the Kardar-Parisi-Zhang (KPZ) 
interface as one tunes the shape of the
noise distribution. We also agree that,
given the height distribution in their Fig.1 (for $d=2$ and $\alpha
=-1/2$) has yet to reach an asymptotic form, there indeed appear to be
long crossover times in the system for low values of
$\alpha $. As we stressed in our original Letter\cite{lett}, the numerical
data we obtained is `strongly suggestive' of non-universal
behaviour, but further numerical work would be required to
clarify the role of temporal crossover. In \cite{lett} we tested for
corrections to scaling (CTS) in the interface width $W(t)$, 
but found them to be small,
namely, the CTS exponent has roughly half the value of
the width exponent $\beta$. 
From Fig. 1 of ref.\cite{pc},
one can see that the height distribution $P$ is more sensitive
to CTS than is $W(t)$, which is the
second moment of $P$. 

It is certainly interesting that the interface width should be relatively
insensitive to CTS, whilst the height distribution is
still crossing over to its asymptotic form. Whether one can
infer from this that $\beta $ is very
slowly crossing over from a measured value of $\simeq 0.13$ to
the `expected' value of $\simeq 0.24$ is not entirely clear.
Although one can obtain very good data collapse for the height
distribution when the noise is Gaussian, the failure
of such a collapse for low values of $\alpha $ may also be 
indicative of more complicated scaling (see point 2 below).
Furthermore, it seems unlikely to us that the percolative
effect has  any bearing on the KPZ physics for $\alpha \ge 0$.
Our numerical data shows that the exponents are very sensitive
to the noise distribution for both positive and negative 
values of $\alpha $. 
Given the difficulties of proving universality from numerical
simulation, it is worthwhile to consider the following two
facts.

1) It is known\cite{nb} that spatially discretized forms of the 
KPZ equation are generally unstable 
(for large coupling) and therefore lie outside
the putative KPZ universality class. Such sensitivity 
to microscopic details is not a property one
normally associates with universality. 

2) It can be shown \cite{en,new} that the deterministic version
of this problem, often referred to as the 
Burgers equation with
random initial data, is sensitive to the shape of
the distribution of initial conditions. If one defines
the initial distribution of the height
(which corresponds to the velocity potential in the
Burgers equation) to be
parameterized by $\alpha $, just as in the present
discussion of the noise distribution (see Eq.(2) of
ref.\cite{pc}), one finds that
the dynamic length scale $L(t)$ increases in time
with an exponent equal to $(1+\alpha)/(d+2+2\alpha)$. 
For a Gaussian initial distribution, $L(t) \sim t^{1/2}$
up to logarithmic corrections.
Not only does this system have non-universal exponents, but it 
has also been shown\cite{new} that naive scaling breaks down due to the
existence of {\it two} important dynamic length scales (corresponding
to $L(t)$ and a diffusive length scale $l_{D} \sim t^{1/2}$).

As a final point, it may be useful to explore new
properties of the KPZ equation in order to gain
much needed insight into scaling and the existence
or otherwise of universality. Prime candidates for
a numerical investigation are persistence 
probabilities\cite{krug}
and the distribution of sign-times\cite{dst}. The former
has recently been studied for the $d=1$ KPZ
equation\cite{kk}, with interesting effects noted as the
shape of the noise distribution is changed. The 
latter is currently under investigation for a
wider range of interface models\cite{ntd}.

Ultimately, the question of the existence of universality in
KPZ physics can only be convincingly answered from a
renormalization group (RG) analysis. As is well appreciated, given
the strong coupling properties of the KPZ equation, such
an analysis is beyond our present expertise. However, a recent
RG calculation\cite{ftj} suggests that the strong coupling regime
may be more intricate than was otherwise imagined. In our opinion,
a clear understanding of the KPZ equation remains a challenge
for the future.

\bigskip
{\obeylines
\noindent T. J. Newman
Department of Physics
Virginia Tech
Blacksburg, VA 24061, USA
}

\medskip

{\obeylines
\noindent M. R. Swift
Department of Physics and Astronomy
University of Manchester
Manchester M13 9PL, UK
}

\medskip

\noindent
PACS numbers: 05.40.+j

\end{document}